# Programmable Nanowrinkle-Induced Room-Temperature Exciton Localization in Monolayer WSe$_2$

Emanuil S. Yanev[1], Thomas P. Darlington[1], Sophia A. Ladyzhets[1], Matthew C. Strasbourg[2], Song Liu[1], Daniel A. Rhodes[3], Kobi Hall[1], Aditya Sinha[1], Nicholas J. Borys[2], James C. Hone[1], and P. James Schuck[1*]

[1]Department of Mechanical Engineering, Columbia University, New York, NY, USA

[2]Department of Physics, Montana State University, Bozeman, MT, USA

[3]Department of Materials Science and Engineering, University of Wisconsin-Madison, Madison, WI, USA

[*]p.j.schuck@columbia.edu

## Abstract:

Localized states in two-dimensional (2D) transition metal dichalcogenides (TMDCs) have been the subject of intense study, driven by potential applications in quantum information science. Despite the rapidly growing knowledge surrounding these emitters, their microscopic nature is still not fully understood, limiting their production and application. Motivated by this challenge, and by recent theoretical and experimental evidence showing that nanowrinkles generate localized room-temperature emitters, we demonstrate a method to intentionally induce wrinkles with collections of stressors, showing that long-range wrinkle direction and position are controllable with patterned array design. Nano-photoluminescence (nano-PL) imaging combined with detailed strain modeling based on measured wrinkle topography establishes a correlation between wrinkle properties, particularly shear strain, and localized exciton emission. Beyond the array-induced super-wrinkles, nano-PL spatial maps further reveal that the strain environment around individual stressors is heterogeneous due to the presence of fine wrinkles that are less deterministic. Detailed nanoscale hyperspectral images uncover a wide range of low-energy emission peaks originating from these fine wrinkles, and show that the states can be tightly confined to regions < 10 nm, even in ambient conditions. These results establish a promising potential route towards realizing room temperature quantum emission in 2D TMDC systems.

## Introduction:

Since their discovery in 2015[1–6], single photon emitters (SPEs) in monolayer transition metal dichalcogenides (TMDC)s have garnered significant interest from the scientific community for a wide range of applications in emerging quantum technologies[7,8]. While substantial progress has been made towards improving both the positioning and purity of SPEs, their underlying nature remains unresolved. Researchers have shown that strain often plays a key role in activating quantum emitters in these materials at low temperatures, and is particularly effective in funneling and localizing excitons[9–16]. More recently, atomistic simulations have predicted that nanoscale wrinkles with extremely sharp radii of curvature can provide localization potentials large enough to create quantum-dot-like states even at room temperature[17,18]. Extensive experimental efforts have been directed at applying local strain to 2D sheets by, for example, using holes[5], pillars[19–26], particles[27,28], bubbles[18,29–31], indentations[32], tips[33], compressible polymers[34], pyramids on cantilevers[35], gaps[36–41], and edges[42–44] as passive and active stressors, all of which have enabled the quasi-deterministic positioning (sub-micron) of local emitting states at cryogenic temperatures. However, the properties and effects of the wrinkles induced by such structures have not been well studied, largely due to the lack of spatial resolution necessary to resolve them.

Here, we show that the engineered properties of stressor array lattices provide an additional knob for controlling exciton localization. By exploiting the fact that wrinkles preferentially form when a 2D layer conforms to non-planar topography, we show that stressor array symmetry guides the dominant orientation of the wrinkles. Using far-field and near-field photoluminescence (nano-PL), we map out the optoelectronic properties of the induced wrinkles from micro to nano length scales, showing enhanced low-energy emission from these areas. Specifically, the localized room-temperature emission coincides with fine nanowrinkles in monolayer (1L) tungsten diselenide ($WSe_2$) sheets emanating from the nanocone stressors. Correlating the nano-PL hyperspectral maps with nanoscale strain maps of the same wrinkles, which are obtained from our AFM topography-based strain modeling, reveals correlations between localized exciton emission and the shear component of strain. Coupled with the greatly reduced defect density in the flux-grown[45] material used, these results provide strong evidence that nanowrinkles are critical for quantum emission in strain-based TMDC systems.

## Results and discussion:

Using lithographically defined topographical features to induce strain in 2D materials has become commonplace[7]. Here, we utilize plasmonic nanocones with tunable tip radii (SI Fig. 1) as our local stressors. This stressor design is motivated by two considerations: i) previous work has linked the observed emission of quantum-dot-like emitters in TMDs at elevated temperatures with local plasmonic enhancement[18,22,23]; and ii) for patterned pillar structures, deeply-confined emission tends to be observed around their rims or edges[43] rather than at a specific point. The cones used in this work increase localization by bending the 2D material about a much smaller point, thereby raising the magnitude of induced strain and facilitating the formation of nanowrinkles possessing sharp curvatures (Fig. 1).

Examples of nanocone-induced wrinkling in 1L-WSe$_2$ can be seen in the tilted scanning electron microscope (SEM) images of Fig. 1b. The 1L-WSe$_2$ conforms strongly to these cones without obvious signs of puncturing. The nanocone arrays were fabricated from substrate-supported Au films through a sequence of lithography, Al$_2$O$_3$ deposition, and ion milling, as illustrated in Fig. 1c. Exfoliated monolayer flakes of WSe$_2$ were then transferred onto the nanocones using polycaprolactone (PCL) on a polydimethylsiloxane (PDMS) stamp. A more in-depth discussion can be found in the methods section.

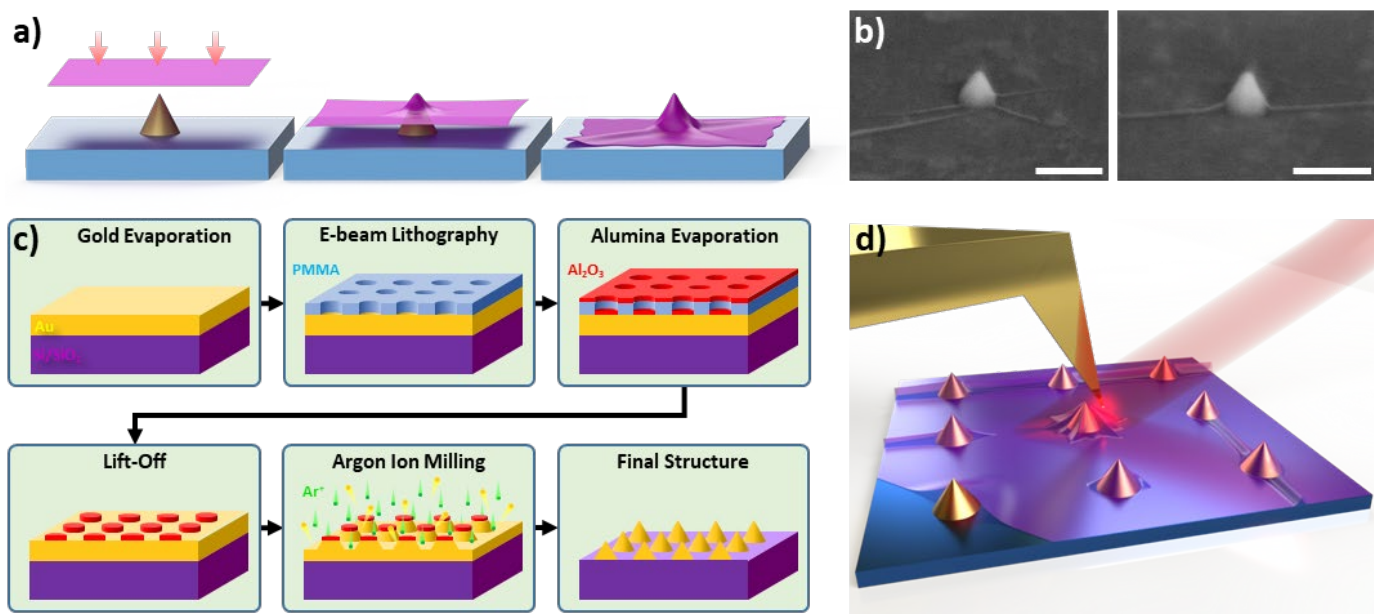

*Figure 1.* **2D materials on nanocones.** (a) Cartoon of wrinkle formation when a sheet conforms to a conical feature. (b) SEM micrographs of wrinkles in monolayer WSe$_2$ on top of isolated nanocones. Scale bars are 200 nm. (c) Fabrication flow for the production of nanocone arrays. (d) Near-field measurement illustration (not to scale).

SEM micrographs of isolated cones reveal the narrow widths of some wrinkles, while also clearly showing that wrinkles can extend for hundreds of nanometers from their point of origin. To investigate the effect that introducing a periodic array of cones has on wrinkle formation, 1L-WSe$_2$ was transferred onto cone arrays with a pitch of 500 nm, in both square and triangular arrangements. AFM (Fig. 2) and optical (Fig. 3) imaging reveals extended networks of interconnected wrinkles bridging many of the lattice sites (Fig. 2a,d), which we term "array wrinkles." Differences in the likelihood of forming these connections are observed between the samples, as shown by the probability distributions in SI Fig. 2. Beyond the array symmetry, we speculate that several aspects influence the formation and direction of such array wrinkles, including the cone lattice spacing, cone height, uncontrolled residual strain from the WSe$_2$ transfer process, and alignment of the array symmetry relative to the crystallographic symmetry of the material; more in-depth investigations will be required for quantifying the role of each of these factors independently and is beyond the scope of the present study.

A careful tally of the wrinkles and their angular orientation illustrates that wrinkles are indeed guided by the chosen array symmetry (Fig. 2c,f). For the square lattice, the predominant directions are vertical and horizontal, corresponding to wrinkle formation between nearest-neighbor sites, as previously shown in graphene[46,47]. Although less common, diagonal wrinkles between second-nearest-neighbors are also present at elevated levels (Fig. 2c; highlighted in blue). In the case of our triangular array, we find that the 60°/240° direction is favored, as is the 0°/180° direction to a lesser extent. Interestingly, we observe almost no array wrinkling along 120°/300°. We attribute this apparent imbalance to the shape of the flake, which is long and narrow, as well as to directional strain unintentionally imparted during the transfer process. We note that while only wrinkles in the flat monolayer area outlined in purple (Fig. 2d) were considered for the histogram, the adjoining bilayer region extending some distance to the left contains a significant number of wrinkles in the 120°/300° direction. Therefore, we posit that a larger monolayer with more internal area will be less influenced by edge effects, and the resulting wrinkles will be more evenly distributed along the primary lattice directions. More generally, our results show that the array degree of freedom can prove to be a useful tuning knob for strain-based 2D systems.

High resolution AFM imaging also allows us to observe finer wrinkle-related features. Specifically, good conformal coverage can give rise to wrinkles that spiral and twist as the sheet contorts around the cone (Fig. 2b,e, center). Detailed scans of these structures reveal wrinkles with out-of-plane radii on the order of the AFM tip (nominally 2 nm) and in-plane meandering curves of 5–10 nm (SI Fig. 3). Such bends may further increase spatial confinement for optical excitations. Additionally, rather than complete conformal coating, in some cases a larger wrinkle (Fig. 2e, top left) or tent-like structure (Fig. 2b, bottom left) is formed, often containing many smaller creases. The optical properties of such nanoscale wrinkles are highlighted below (Figs. 4, 5).

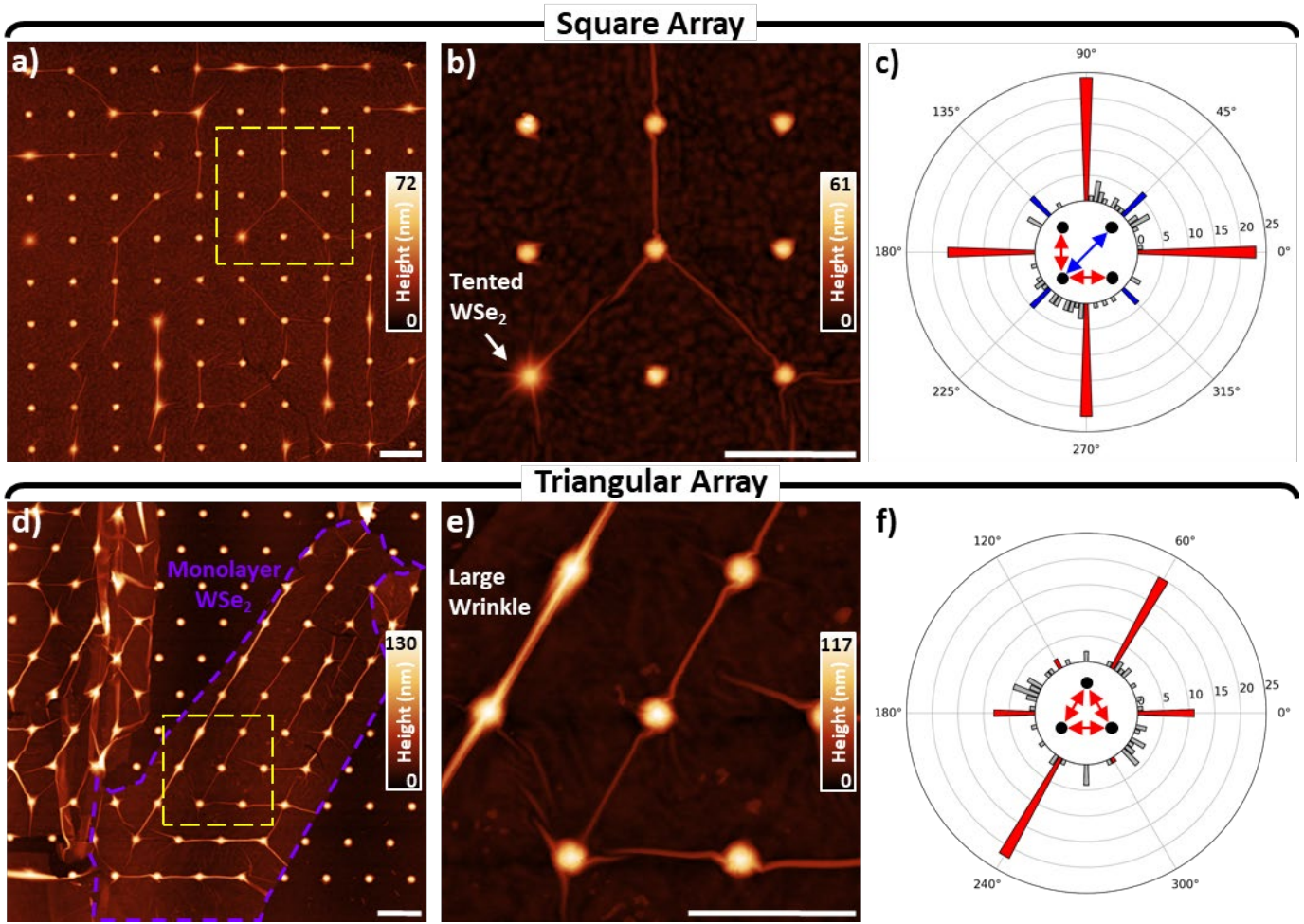

*Figure 2.* **AFM characterization of array wrinkles on substrates with different lattice symmetries.** (a) Wrinkles in a monolayer of $WSe_2$ on a square array of nanocones. (b) Detail of yellow square in (a) exhibiting both strong and weak conformity of the $WSe_2$ to the cone. (c) A polar histogram of wrinkle directions in (a). (d) $WSe_2$ on a triangular array of nanocones, with the monolayer region outlined in purple. (e) Detail of yellow square in (d) showing kinking of some wrinkles as they spiral around cones. (f) A polar histogram of wrinkle directions in (d). Only wrinkles in the monolayer region were counted. All scale bars are 500 nm.

To investigate the exciton emission properties of wrinkles induced by the nanocone stressor arrays, we first performed far-field hyperspectral PL mapping on the square array sample using 633 nm CW excitation. Topography of the entire scanned region is shown with an inverted and truncated color scale to highlight the small monolayer wrinkles (Fig. 3a) relative to the much larger wrinkles formed in the multilayer region, which are visible even with a diffraction-limited white-light microscope (SI Fig. 4). Typical far-field spectra exhibit the standard primary exciton (PX) peak[18] (Fig. 3b) that tends to increase in intensity near wrinkles. Additionally, maps of PL intensity vary as a function of emission wavelength. Images of intensity generated by sweeping a 30 nm wide integration window from 740 nm to 830 nm show significant spatial inhomogeneity, particularly when the integration window is positioned on the low-energy tail of the PX spectrum. Example locations that exhibit strong spatial-spectral inhomogeneity on the filtered hyperspectral maps are highlighted in yellow in Fig. 3c-e. The spatial variations suggest the existence of localized low-energy states and motivate further investigation with nano-optical techniques[18,48].

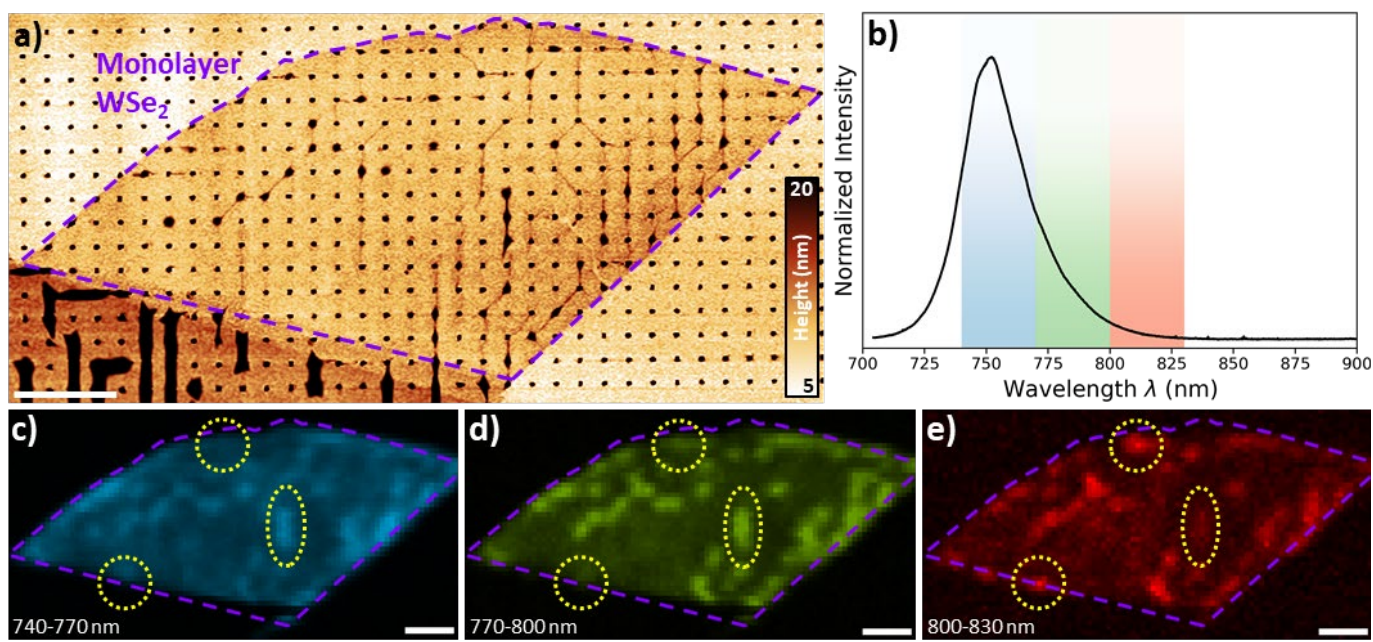

*Figure 3.* **Heterogeneous far-field optical properties.** (a) AFM topography of the confocal PL scan area. The dark lines are wrinkles between the cones. (b) Averaged spectrum from the entire region. (c-e) Confocal PL maps showing the integrated intensity in the three correspondingly colored spectral windows in (b). The areas circled in yellow highlight the presence of localized states at different energies. All scale bars are 2 μm.

Focusing on a wrinkled region around the center of the monolayer (see rightmost yellow region in Fig. 3d,e), we collected hyperspectral nano-PL scans following the protocol detailed in methods. The AFM topography map (Fig. 4a) reveals many wrinkles in the area, and enhanced exciton emission over all wavelengths is seen primarily from the area with the larger wrinkle near the middle (Fig. 4b,c). We note that the flat regions are dim, due primarily to PL quenching by a thin layer of Au remaining on the sample surface, which serves to suppress unwanted background signal from 1L-WSe$_2$ in these areas. In the large wrinkle, the WSe$_2$ is sufficiently separated from the substrate to overcome the quenching. Nano-PL emission maps are shown for the spectral window encompassing the (unstrained) PX peak (Fig. 4b, 740-770 nm) and for a lower-energy spectral window (Fig. 4c, 770-900 nm). In the low-energy window, emission is most intense in the wrinkle and in regions immediately surrounding the cones (Fig. 4c). In the higher energy window, while emission from the topmost cone apex is relatively dim, sizeable emission is present at the apexes of the other two cones (Fig. 4b). Rather than being fully quenched, Purcell-enhanced emission at the apexes of plasmonic nanocone antennas can shift the balance between radiative exciton recombination, nonradiative quenching, and funneling to lower-energy states, resulting in detectable PX PL signal arising from such regions[9–16,49–52]. We note that cone-to-cone variations in enhancement and quenching are to be expected given nanoscale fabrication-related substrate heterogeneity.

Evidence of localized low-energy exciton emission from different areas within the wrinkled region can be seen in the five sample spectra shown in Fig. 4d, corresponding to the locations marked with white arrows in Fig. 4c. These representative spectra illustrate the wide variety of emission lineshapes observed throughout the sample and infer the presence of a rich, heterogeneous, nanoscale landscape of strain imparted by topographical stressors.

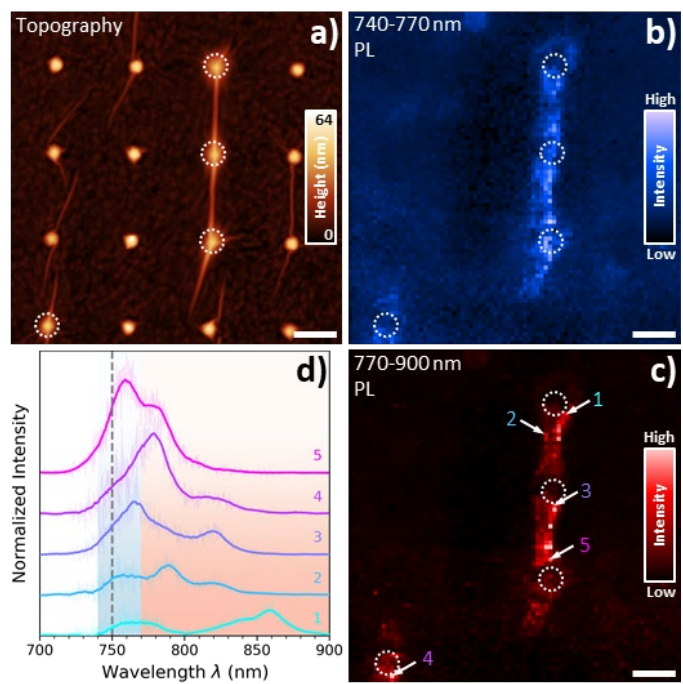

*Figure 4*. **Near-field optical investigation of a large array wrinkle.** (a) AFM topography of the nano-PL scanned region. White circles are a guide for the eye, marking the approximate location of the nanocone stressors. (b) Near-field intensity in the primary exciton spectral range of 740-770 nm. (c) Near-field intensity in the low-energy spectral range of 770-900 nm. (d) Select spectra, taken from the locations marked with white arrows in (c), showing a diverse assortment of low-energy states. The blue and red colored regions correspond to the integration ranges for (b) and (c), respectively. The bold solid lines are the result of a moving average overlaid on top of the raw data for clarity. All scale bars are 250 nm.

Zooming in on the cone at the top of the array wrinkle that is highlighted in Fig. 4 reveals multiple small wrinkles or creases in the tent-like portion of the monolayer (Fig. 5a; as noted above, less PX emission is observed from this region). Several of these fine wrinkles are explicitly labeled to differentiate them from the larger array wrinkle that forms between neighboring cones. The dashed circles in Fig. 5a-f demarcate the base of the cone underneath the tented WSe$_2$. Integrating the intensity over the full spectral range clearly shows bright nano-PL at the cone periphery (Fig. 5b), as previously discussed. In Fig. 5c, the spectral median was calculated for all pixels above a minimum threshold intensity. A large shift of ~80 nm (~160 meV) corresponds to the most intense regions of nano-PL. More generally, a comparison of panels (c) and (d) shows a clear correlation between emission energy and intensity within the strain-localized regions. This correlation is further quantified in SI Fig. 5. More direct comparison with strain is achieved by modeling local strain fields using a technique[53] based on AFM topography previously developed for nanobubbles. Generalizing the method to include wrinkles, we were able to estimate the normal (panel e) and shear (panel f) components of strain in the vicinity of the cone. These calculations reveal fairly large strains of ~2.5%, and that the brightest LX emission comes from regions with high shear.

Multiple spectra of interest were extracted from the 10 numbered pixels in panel (b) and plotted in Fig. 5g. From these spectra, we observe that LX emission can span a broad range of energies due to the complex, heterogeneous strain environment around a single stressor, with multiple local maxima in strain serving to confine excitons in distinct but adjacent regions. The same behavior is observed in fine wrinkles around other cone locations, with LX peaks occurring at even longer wavelengths—out to, and potentially beyond, 900 nm (SI Fig. 6; long-wavelength detection is limited by detector bandwidth in our setup). Panel (h) shows 5 neighboring spectra originating from a line cut through one of the fine wrinkles. A prominent LX peak is only observed from the single 10 nm pixel at the middle of the line cut. This demonstrates the high degree of spatial confinement of excitonic states by the fine wrinkles, in good agreement with prior work[17,18]. Similar confinement can be seen at other locations in this data set (SI Fig. 7a,b), and others (SI Fig. 7c,d). These results highlight the potential for substrate-induced nanowrinkles as sources of room-temperature localized exciton state emission, as well as the challenges in deterministically positioning such states with nanoscale precision.

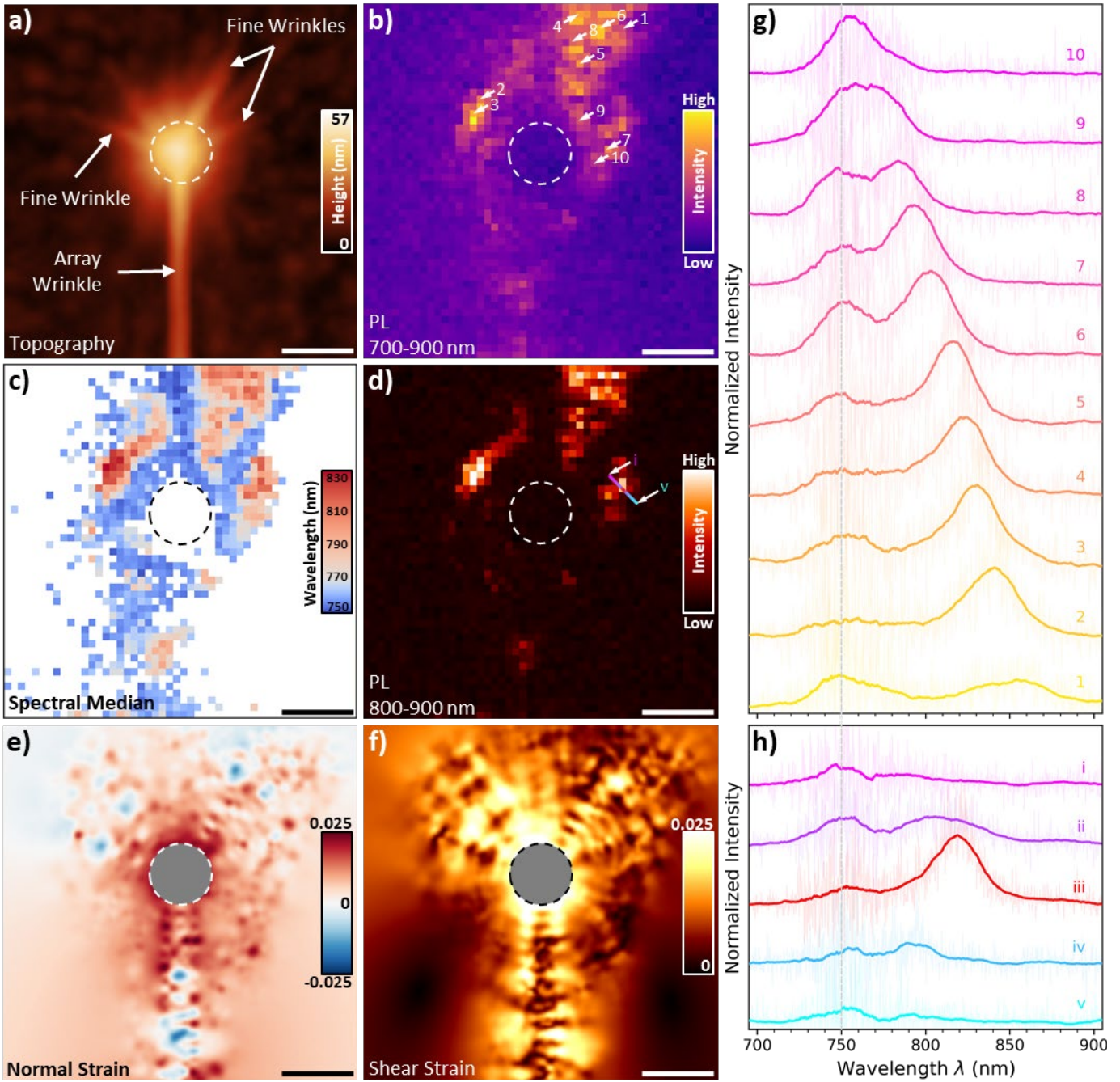

*Figure 5. Near-field optical properties of fine nanowrinkles.* (a) AFM topography of the nano-PL scan region showing very small wrinkles radiating out from the top of the cone. (b) Nano-PL integrated intensity from 700–900 nm. (c) A spatial map of the spectral median, where pixels with an integrated intensity below some threshold value were excluded for clarity. (d) Nano-PL from localized states in the range of 800–900 nm. (e) Normal and (f) shear strains calculated from AFM topography exhibit maximum values of ~2.5% away from the cone center, which has been excluded from the images due to ambiguities in the strain state at the tip apexes and to highlight the fine wrinkles.

*(g) Spectra taken from the locations marked with white arrows in (b), clearly showing that LX states span a spectral range of more than 100 nm below the primary exciton, presumably due to varying amounts of strain. (h) Spectra taken from a cut through one of the fine wrinkles. The location is marked with a colored line in panel (d), and the start and finish are labeled with lowercase roman numerals i and v, respectively. The LX peak is well confined to a single 10 nm pixel in the middle. All scale bars are 100 nm.*

## Conclusions:

In summary, placing 2D materials on top of nanocone substrates is an effective way to create strained wrinkles with preferential orientations prescribed by the symmetry of the patterned features. In addition to array wrinkles that connect neighboring cones, fine wrinkles are formed by topographic stressors. Although their shapes and directions are more stochastic, their ability to strongly confine excitons is much greater, potentially leading to SPEs at room temperature. Using hyperspectral nano-PL imaging, we observe a wide gamut of low-energy peaks—spanning a range of over 235 meV—around a single nanocone stressor. This highlights the fact that the strain environment around artificial structures is often highly inhomogeneous and may need to be better controlled for future applications that seek uniform emitter properties. The emission from these states is extremely localized in some cases, reaching resolution-limited confinement in scans with step sizes of 2-10 nm. Given their spatial extent at room temperature, it is possible that these sites could exhibit single photon emission, as previously suggested by theory[17]. This work adds a new element of control while also elucidating the role of underlying nanoscale heterogeneity, yielding a deeper understanding of strain on localized exciton behavior, which will be crucial for deterministically positioning and tuning localized emitters for nano- and quantum-photonics applications.

# Methods:

## Sample fabrication

Arrays of nanoscale gold cones were fabricated through a process adapted from Schäfer et al[54]. First, a 5 nm thick titanium adhesion layer was evaporated onto a Si/SiO$_2$ substrate, followed by 100 nm of gold. The thickness of these layers determines the maximum height of the final structures. PMMA and MMA copolymer were spun on top of the gold to form a bilayer electron-beam resist stack of around 110 nm thick. This resist layer was subsequently exposed and developed with a mixture of IPA and MIBK to form an array of circular wells. Aluminum oxide was then evaporated through the openings, leaving behind thin disks on the gold surface after lift-off in acetone. Finally, argon ion milling was performed at normal incidence until the aluminum oxide hard mask was fully consumed. Due to the angular dependence of both the sputtering yield and sidewall redeposition[55], this process produces features with sloped edges. Since circular masks were used, the result after milling was an array of cones with tip radii as low as 2 nm in optimized cases.

Flakes of 2D materials were then dry transferred onto the cone arrays. High quality TMDC crystals were first grown in house via the flux synthesis method[45], and then mechanically exfoliated onto Si/SiO$_2$ substrates using the well-known Scotch tape technique[56]. Monolayers were then directly picked up using a PCL stamp. Suhan Son et al. provide a detailed discussion of PCL stamp preparation and usage[57]. In short, PCL was spun to a uniform thickness and placed on top of a PDMS square attached to a glass slide, in similar fashion to stacking methods with other polymers. Flakes were brought into contact with the PCL by heating the substrate from 50° C to 59° C, and then picked up by cooling down to 40° C. The monolayers were then transferred onto the nanocone substrates by increasing the temperature above PCL's melting point of 60° C and slowly lifting the stamp. Finally, the polymer was removed by submerging the samples in hot (70-80° C) acetone for one hour.

## Optical measurements:

Room-temperature PL was collected from the nanocone samples using a near-field scanning optical microscope (OmegaScope; Horiba Scientific) in conjunction with a Raman spectrometer (LabRAM HR Evolution; Horiba Scientific). Light from a 633 nm excitation laser was passed through a microscope objective (100x, 0.7 NA) at an angle of 65° relative to the surface normal and focused onto the apex of a gold coated AFM probe (AppNano Omni™ TERS-NC-Au). Collection of the sample luminescence was also accomplished with the same objective. Regular non-contact AFM scans were performed to locate the regions of interest, after which a special scanning modality (Dual-Spec; Horiba Scientific) was used to collect the PL. In this mode, the tip is brought into and out of contact with the sample at each pixel and a spectrum is collected at both tip positions. Due to a strong distance dependence, the spectra collected with the tip retracted away from the surface are almost entirely devoid of near-field contributions and serve as a good approximation of the far-field background. These were subsequently subtracted from the spectra obtained with the tip in contact to produce the hyperspectral nano-PL maps in Figs. 4 and 5, and SI Figs. 6 and 7.

## Finite Element Strain Modeling:

The strain was calculated following the method described in reference [53] which models the monolayer TMD as a classical plate. The Airy stress tensor is first calculated by solving the 2nd of the Foppl-von Karman equation using the Simple Finite Element Methods in Python[58]. The strain is then estimated from the Airy stress function via numerical differentiation.

## Acknowledgements:

This work was supported by the National Science Foundation through award NSF DMR No. 2004437. T.P.D., J.C.H. and P.J.S. acknowledge support from Programmable Quantum Materials, an Energy Frontier Research Center funded by the US Department of Energy (DOE), Office of Science, Basic Energy Sciences (BES), under award DE-SC0019443. Nanostressor fabrication was partially supported through a US Department of Energy, Office of Science Graduate Student Research (SCGSR) award (E.Y.) and Honda Research Institute USA, Inc. N.J.B. also acknowledges the MonArk NSF Quantum Foundry supported by the National Science Foundation Q-AMASE-i program under NSF award No. DMR-1906383.